\documentclass[a4paper,11pt]{article}

\usepackage{amsmath}
\usepackage{bm}
\usepackage{siunitx}
\usepackage{graphicx}
\usepackage{hyperref}
\usepackage{amsmath,amsthm,amssymb}
\usepackage{xcolor}
\usepackage{epstopdf}
\usepackage{overpic}
\usepackage{bbold}
\usepackage{textcomp}
\usepackage{physics}

\usepackage{authblk}

\providecommand{\keywords}[1]
{
  \small
  \textbf{\textit{Keywords---}} #1
}

\begin{document}
\title{Proposal for a micromagnetic standard problem: domain wall pinning at phase boundaries}

\author[1]{Paul Heistracher \thanks{paul.thomas.heistracher@univie.ac.at}}
\author[1]{Claas Abert}
\author[1]{Florian Bruckner}
\author[2,3]{Thomas Schrefl}
\author[1]{Dieter Suess}

\affil[1]{University of Vienna, Faculty of Physics, Physics of Functional Materials, Kolingasse 14-16, 1090,  Vienna, Austria}
\affil[2]{Christian Doppler Laboratory for Magnet design through physics informed machine learning, Viktor Kaplan-Straße 2E, 2700 Wiener Neustadt, Austria}
\affil[3]{Department for Integrated Sensor Systems, Danube University Krems, Viktor Kaplan-Straße 2E, 2700 Wiener
	Neustadt, Austria}
\maketitle

\begin{abstract}
We propose a novel micromagnetic standard problem calculating the coercive field for unpinning a domain wall at the interface of a multiphase magnet.
This problem is sensitive to discontinuities in material parameters for the exchange interaction, the uniaxial anisotropy, and the spontaneous magnetization.
We derive an explicit treatment of jump conditions at material interfaces for the exchange interaction in the finite-difference discretization.
The micromagnetic simulation results are compared with analytical solutions and show good agreement.
The proposed standard problem is well-suited to test the implementation of both finite-difference and finite-element simulation codes.
\end{abstract}


\section{Introduction}
Micromagnetic simulations have proven to be an essential tool to support the research and development of magnetic devices such as magnetic random access memory \cite{duine_synthetic_2018} or magnetic sensors \cite{suess_topologically_2018}.
Through ongoing miniaturization, these devices enter the domain of micromagnetics, where sub-micrometer magnetic structures such as domain walls or vortices become relevant.

To validate the implementation of such micromagnetic simulation codes and prevent common pitfalls, the micromagnetics community proposes specific test cases referred to as \textit{standard problems}.
One of the most prominent collections of such standard problems for ferromagnetic materials is from the micromagnetic modeling activity group ($\mu$MAG) from the National Institute of Standards and Technology.
This collection is hosted on their website \footnote{\url{https://www.ctcms.nist.gov/~rdm/mumag.org.html}}.
These standard problems test magnetic properties resulting from the modeled micromagnetic interactions like hysteresis loops (sp1, sp2), static behavior (sp3), dynamics of a magnetic film (sp4), and spin-transfer torque (sp5, \cite{najafi_proposal_2009}).
Further standard problems are proposed in more recent literature. These include numerical examples on spin waves \cite{venkat_proposal_2013}, ferromagnetic resonance \cite{baker_proposal_2017}, and the Dzyaloshinskii–Moriya interaction \cite{cortes-ortuno_proposal_2018}.
These standard problems have become an essential part of the validation of micromagnetic simulation software for the micromagnetic community.
However, all of these standard problems treat a single magnetic material.
This manuscript, therefore, proposes a micromagnetic standard problem assessing the correct treatment of interface effects in a multiphase magnet.

\section{Micromagnetic model}
The micromagnetic model is a semi-classical description of magnetism as it combines quantum-mechanical effects, such as the exchange interaction, with a classical continuous field description of magnetism \cite{abert_micromagnetics_2019}.
The central assumption of the micromagnetic model is that ferromagnetic ordering due to the exchange interaction dominates the magnetic ordering on a local scale.
This ferromagnetic ordering keeps the magnetization in parallel on a characteristic length scale $\lambda$, which is well above the lattice constant $a$ of the material.
For distinct magnetic moments $\bm{S_i}$ and $\bm{S_j}$ at locations $\bm{r_i}$ and $\bm{r_j}$, respectively, we can assume
\begin{equation}
\bm{S_i} \approx \bm{S_j} \ \ \  \text{for} \ \ \ |\bm{r_i} - \bm{r_j}| < \lambda \gg a.
\end{equation}
The strong ferromagnetic ordering on a local scale gives rise to the \textit{continuum approximation}, where we introduce a continuous vector field $\bm{M}(\bm{r})$ which approximates the local spin density.
Given a homogeneous density of elementary spins, we can express the magnetization in terms of a unit-vector field $\bm{m}(\bm{r})$ with
\begin{equation}
	\bm{M}(\bm{r}) = M_\text{s} \bm{m}(\bm{r}) \ \ \ \text{with} \ \ \ |\bm{m}(\bm{r})| = 1,
\end{equation}
where $M_\text{s}$ is the spontaneous magnetization in \SI{}{J/T/m^3}.

\subsection{Energy contributions}
The micromagnetic energy contributions considered in the proposed standard problem are the exchange interaction $E^\text{ex}$, uniaxial anisotropy $E^\text{ani}$, and external field contributions $E^\text{ext}$.
Therefore, the combined energy $E$ of the system reads
\begin{equation}
E = E^\text{ex} + E^\text{ani} + E^\text{ext}.
\end{equation}
We neglect micromagnetic demagnetization effects in the scope of this manuscript in order to preserve a simple analytical solution.
Analytic solutions of interface problems including magnetostatic effects are discussed in \cite{kronmuller_theory_1987} and \cite{kronmuller_pinning_2008}.

With an expression of the accumulated energy terms $E$, we can calculate the effective field $\bm{H}^\text{eff}$ as the functional derivative of $E$ with respect to the magnetization $\bm{m}$,
\begin{equation}\label{eq:Heff_via_dE}
- \mu_0 M_\text{s} \bm{H}^\text{eff} = \frac{\delta E}{\delta \bm{m}},
\end{equation}
where $\mu_0 = \SI{1.256 637 062 12 e-6}{N/A^2}$ is the vacuum magnetic permeability.
This effective field is then used to describe the dynamics of the system, as described in section \ref{sec:LLG}.

\subsubsection{Exchange interaction}
In ferromagnetic materials, the quantum mechanical exchange interaction causes the elementary spins to prefer a parallel alignment.
In the micromagnetic model, this exchange interaction is accounted for by a phenomenological continuum description with an energy contribution $E^{\text{ex}}$ which reads
\begin{equation}
E^{\text{ex}}
= \int_{\Omega} A (\grad{\bm{m}})^2 \ \text{d}\bm{x},
\end{equation}
where $A$ is the micromagnetic exchange constant in $\SI{}{J/m}$ and $\Omega$ is the magnetic domain.

The differential $\delta E^{\text{ex}}$ can be calculated by variational calculus \cite{abert_micromagnetics_2019} applying an arbitrary test function $\bm{v} \in V$ with $V$ as the function space of the magnetization $\bm{m}$
\begin{equation}\label{eq:Eex}
\begin{split}
\delta E^{\text{ex}} (\bm{m}, \bm{v}) 
& = \frac{\text{d}}{\text{d} \epsilon} \Big[ \int_{\Omega} A [\grad{(\bm{m} + \epsilon \bm{v})}]^2 \ \text{d}\bm{x} \Big]_{\epsilon = 0}
\\ & = 2 \int_{\Omega} A \grad{\bm{m}} : \grad{\bm{v}} \ \text{d}\bm{x}
\\ & = - 2 \int_{\Omega} [\grad{} \cdot (A \grad{\bm{m}})] \cdot \bm{v} \ \text{d}\bm{x} + 2 \int_{\partial\Omega} A \partial_{\bm{n}} \bm{m} \cdot \bm{v} \ \text{d}\bm{s}
,
\end{split}
\end{equation}
where in the first step, we expanded the square brackets and applied the derivative with respect to $\epsilon$.
The operator "$:$" denotes the sum of the Hadarmard product $a:b = \sum_{i,j} a_{ij} b_{i,j}$.
In the second step, we applied partial integration and used the divergence theorem, obtaining a surface integral.
In order to apply the divergence theorem, we assume that $A \grad{\bm{m}}$ is continuously differentiable, i.e., it does not exhibit any jumps.
The expression $\partial_{\bm{n}} \bm{m} := \frac{\partial \bm{m}}{\partial \bm{n}}$ denotes the directional derivative with respect to the surface normal vector $\bm{n}$.
Using equation (\ref{eq:Heff_via_dE}), we obtain the effective field for the exchange interaction

\begin{equation}\label{eq:Hex}
\bm{H}^{\text{ex}}
= \frac{-1}{\mu_0 M_\text{s}} \frac{\delta E^{\text{ex}}}{\delta \bm{m}}
= \frac{-1}{\mu_0 M_\text{s}} (-2 \grad{\cdot A \grad{\bm{m}}})
.
\end{equation}
The second term yields the exchange boundary condition
\cite[eq. 63]{abert_micromagnetics_2019}
\begin{equation}\label{eq:m_x_2Admdn}
2 A \ \partial_{\bm{n}} \bm{m} = 0
,
\end{equation}
which demands that the directional derivative disappears at the boundary.

\begin{figure}[h!]
	\centering
	\includegraphics{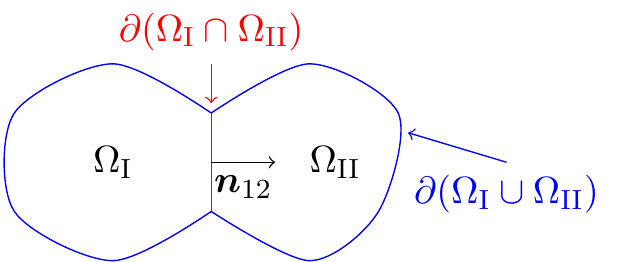}
	\caption{Illustration of a two-phase magnetic meterial consisting of integration regions $\Omega_{\text{I}}$ and $\Omega_{\text{II}}$ with conjunct boundary $\partial (\Omega_{\text{I}} \cap \Omega_{\text{II}})$, disjunct boundary $\partial (\Omega_{\text{I}} \cup \Omega_{\text{II}})$, and normal vector $\bm{n}_{12}$ pointing from $\Omega_{\text{I}}$ to $\Omega_{\text{II}}$ .}
	\label{fig:int_regions}
\end{figure}

With this derivation in mind, we now discuss the case of an inhomogeneous exchange constant $A(\bm{x})$.
We assume a two-phase magnetic material with regions $\Omega_{\text{I}}$ and $\Omega_{\text{II}}$, as illustrated in Figure \ref{fig:int_regions}.
The value of the exchange parameter differs between the two regions and is specified to $A(\bm{x})=A_\text{I}$ in region region $\Omega_{\text{I}}$ and $A(\bm{x})=A_\text{II}$ in region $\Omega_{\text{II}}$.
To calculate the differential $\delta E^{\text{ex}}$, we split the integration domain into the two magnetic phases and perform the steps from equation (\ref{eq:Eex}) for each region separately.
We then combine the integration domain of the separate terms and obtain two additional interface terms for the conjunct boundary.

\begin{equation}\label{eq:Eex_twophase1}
\begin{split}
\delta E^{\text{ex}} (\bm{m}, \bm{v}) 
& = \frac{\text{d}}{\text{d} \epsilon} \Big[
\int_{\Omega_\text{I}} A_\text{I} [\grad{(\bm{m} + \epsilon \bm{v})}]^2 \ \text{d}\bm{x}
+ \int_{\Omega_\text{II}} A_\text{II} [\grad{(\bm{m} + \epsilon \bm{v})}]^2 \ \text{d}\bm{x}
\Big]_{\epsilon = 0}
\\ & = - 2 \int_{\Omega} [\grad{} \cdot (A(\bm{x}) \grad{\bm{m}})] \cdot \bm{v} \ \text{d}\bm{x} + 2 \int_{\partial (\Omega_{\text{I}} \cup \Omega_{\text{II}})} A(\bm{x}) \partial_{\bm{n}} \bm{m} \cdot \bm{v} \ \text{d}\bm{s}
\\ & \ \ \  +  \int_{\partial (\Omega_{\text{I}} \cap \Omega_{\text{II}})} A_\text{I} \partial_{\bm{n}} \bm{m} \cdot \bm{v} \ \text{d}\bm{s}
+  \int_{\partial (\Omega_{\text{I}} \cap \Omega_{\text{II}})} A_\text{II} \partial_{\bm{n}} \bm{m} \cdot \bm{v} \ \text{d}\bm{s}
\end{split}
\end{equation}
Replacing the surface normal vector $\bm{n}$ for each subdomain at the interface between $\Omega_{\text{I}}$ and $\Omega_{\text{II}}$ with $\bm{n}_{12}$,
the normal vector pointing from $\Omega_{\text{I}}$ to $\Omega_{\text{II}}$, we obtain
\begin{equation}\label{eq:Eex_twophase2}
\begin{split}
\delta E^{\text{ex}} (\bm{m}, \bm{v})
& = - 2 \int_{\Omega} [\grad{} \cdot (A(\bm{x}) \grad{\bm{m}})] \cdot \bm{v} \ \text{d}\bm{x} + 2 \int_{\partial (\Omega_{\text{I}} \cup \Omega_{\text{II}})} A(\bm{x}) \partial_{\bm{n}} \bm{m} \cdot \bm{v} \ \text{d}\bm{s}
\\ & \ \ \  +  \int_{\partial (\Omega_{\text{I}} \cap \Omega_{\text{II}})} A_\text{I} \partial_{\bm{n}_{12}} \bm{m} \cdot \bm{v} \ \text{d}\bm{s}
- \int_{\partial (\Omega_{\text{I}} \cap \Omega_{\text{II}})} A_\text{II} \partial_{\bm{n}_{12}} \bm{m} \cdot \bm{v} \ \text{d}\bm{s}
\end{split}
\end{equation}
The first two terms in equation (\ref{eq:Eex_twophase2}) correspond to the result of equation (\ref{eq:Eex}).
The first term yields an effective field, while the second term yields a boundary condition defined over the disjunct boundary $\partial (\Omega_{\text{I}} \cup \Omega_{\text{II}})$.
Compared to the homogeneous case, two additional terms occur for the conjunct boundary $\partial (\Omega_{\text{I}} \cap \Omega_{\text{II}})$ at the interface.
These two terms must vanish for all test functions $\bm{v}$.
Therefore, we obtain the interface condition
\begin{equation}\label{eq:interface_jump_0}
A_\text{I} \partial_{\bm{n}_{12}} \bm{m} = A_\text{II} \partial_{\bm{n}_{12}} \bm{m}
.
\end{equation}
This corresponds to the interface condition in \cite{BrownWilliamFuller1963M} and \cite[eq. 6]{kronmuller_micromagnetic_2002}.
The implications of this jump condition for a finite-difference discretization are discussed in section \ref{sec:exchange_interface_condition}.

\subsubsection{Uniaxial anisotropy and external field}
The second energy contribution we consider is the uniaxial anisotropy energy, which prefers parallel or antiparallel alignment of the magnetization along a specific direction.
With $\bm{e}_{\text{u}}$ being the unit vector along the easy axis, this energy reads
\begin{equation}
E^{\text{ani}} = - \int_{\Omega} K_{\text{u1}}(\bm{m} \cdot \bm{e}_{\text{u}})^2 \ \text{d}\bm{x}
,
\end{equation}
where $K_{\text{u1}}$ is the anisotropy constant in $\SI{}{J/m^3}$.
For this interaction, the resulting effective field reads
\begin{equation}\label{eq:ani_micro}
\bm{H}^{\text{ani}} = \frac{2 K}{\mu_0 M_\text{s}} \bm{e}_\text{u} (\bm{e}_\text{u} \cdot \bm{m})
.
\end{equation}

Furthermore, the presence of an external magnetic field $\bm{H}^{\text{ext}}$ contributes to the energy with

\begin{equation}
E^{\text{ext}} = - \mu_0 \int_{\Omega} M_\text{s} \bm{m} \cdot \bm{H}^{\text{ext}} \ \text{d}\bm{x}
.
\end{equation}

\subsection{Landau-Lifshitz-Gilbert equation}\label{sec:LLG}
We combine the three discussed energy contributions into one effective field $\bm{H}^\text{eff} = \bm{H}^{\text{ex}} + \bm{H}^{\text{ani}} + \bm{H}^{\text{ext}}$.
This effective field can then be used in the Landau-Lifshitz-Gilbert (LLG) equation to describe the time evolution of the magnetization
\begin{equation}
\frac{\partial \bm{m}}{\partial t} = - \frac{\gamma}{1+ \alpha^2} \bm{m} \cross \bm{H}^\text{eff} - \frac{\alpha \gamma}{1+ \alpha^2} \bm{m} \cross \big(\bm{m} \cross \bm{H}^\text{eff}\big)
,
\end{equation}
where $\gamma = \mu_0 |\gamma_e| \approx \SI{2.2128e5}{m/As}$ is the reduced gyromagnetic ratio ($\gamma_e$ being the electron gyromagnetic ratio) and $\alpha \ge 0$ is a dimension-less damping parameter.
The first term in the LLG equation describes spin-precession, the second term energy dissipation.

\section{Field contributions in finite-differences}
The two major techniques to discretize the micromagnetic continuum model are the finite-difference method and the finite-element method.
In the finite-difference discretization, the magnetic domain is discretized into a uniform cuboid mesh, and the differential operators are approximated by finite differences.
We choose the degrees of freedom to be in the center of the simulation cells for both the magnetization field and the material parameters.
This is the predominant choice in literature and is also used by
\cite{m._j._donahue_oommf_1999} and \cite{vansteenkiste_design_2014}.

\subsection{Exchange field: homogeneous magnet}
In the case of a single-phase magnet with a homogeneous exchange constant $A$, the exchange field in equation (\ref{eq:Hex}) can be simplified to the form
\begin{equation}\label{eq:Hex_hom}
\bm{H}_{\bm{i}}^{\text{ex, hom}}
= \frac{2 A}{\mu_0 M_{\text{s},\bm{i}}} \laplacian{\bm{m_i}}
\approx \frac{2 A}{\mu_0 M_{\text{s},\bm{i}}} \sum_{k=x,y,z} \frac{\bm{m}_{\bm{i} + \bm{e}_k} -2 \bm{m_i} + \bm{m}_{\bm{i} - \bm{e}_k} }{\Delta_k^2}
,
\end{equation}
where $\bm{i}=\{i, j, k\}$ is a three-dimensional cell index,
$\bm{e}_x, \bm{e}_y, \bm{e}_z$ are unit vectors along each axis,
and $\Delta_x, \Delta_y, \Delta_z$ are the cell sizes of the finite-difference discretization.
The three-dimensional discrete Laplace operator $\laplacian{}$ is approximated with a second-order central derivative.

Micromagnetic solvers often implement the exchange field for a homogeneous material by means of a convolution.
The sum in equation (\ref{eq:Hex_hom}) can be interpreted as a constant Laplace kernel and applied onto the magnetization.
This property can no longer be used for inhomogeneous, cell-dependent exchange values $A_{\bm{i}}$, as no common kernel exists.
In this case, the effective field can be calculated by a sparse matrix-vector product.

\subsection{Exchange field: inhomogeneous magnet and interface conditions}\label{sec:exchange_interface_condition}

\begin{figure}[h!]
\centering
\includegraphics{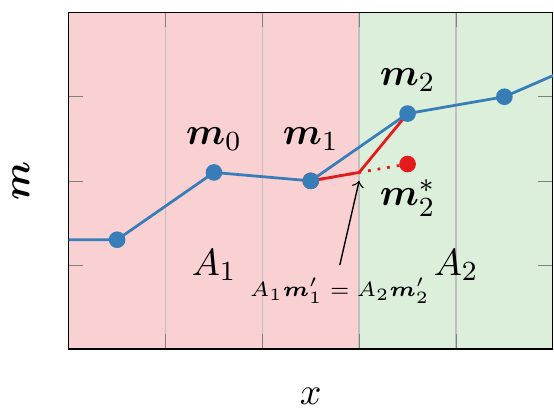}
\caption{Illustration of jump-condition handling at a material interface in finite-differences. Cell-based magnetization values (blue) with virtual virtual magnetization $\bm{m}_2^*$ (red) for evaluating the three-point Laplace stencil at the location of $\bm{m}_1$.}
\label{fig:fd_cells}
\end{figure}

In a multiphase material with different exchange values $A_{\bm{i}}$ per cell, interface conditions as given in equation (\ref{eq:interface_jump_0}) arise and must be treated explicitly when calculating the exchange field, rendering equation (\ref{eq:Hex_hom}) invalid.

In the following, we discuss the handling of the interface condition for the one-dimensional case and then generalize to three dimensions.
Figure (\ref{fig:fd_cells}) shows a cell-centered finite-difference discretization of a two-phase magnet.
The value of the exchange constant
changes form $A_1$ (red) to $A_2$ (green) between the magnetizations $\bm{m}_1$ and $\bm{m}_2$.
At the exchange coupling interface, the first derivative of the magnetization exhibits a jump given by equation (\ref{eq:interface_jump_0}).
For this one-dimensional example, the interface condition becomes
\begin{equation}\label{eq:interface_jump}
A_1 \bm{m}_1^{\prime} = A_2 \bm{m}_2^{\prime}
,
\end{equation}
where $\bm{m}_1^{\prime}$ is the right-hand derivative at the interface and $\bm{m}_2^{\prime}$ is the left-hand derivative, indicated in Figure \ref{fig:fd_cells} by the red line.
This interface condition must be taken into account when considering the discretized Laplace operator.

In first order, we approximate the difference between the two neighboring magnetizations $\bm{m}_1$ and $\bm{m}_2$ with the arithmetic mean of the first derivatives $\bm{m}_1^{\prime}$ and $\bm{m}_2^{\prime}$
\begin{equation}\label{eq:m_2}
\bm{m}_2 = \bm{m}_1 + \frac{\Delta_x}{2} \big(\bm{m}_1^{\prime} + \bm{m}_2^{\prime}\big)
+ \mathcal{O}(\Delta_x^2)
.
\end{equation}
In alternative terms, given the magnetization $\bm{m}_1$ and both derivatives $\bm{m}_1^{\prime}$ and $\bm{m}_2^{\prime}$, the magnetization $\bm{m}_2$ is obtained by starting at $\bm{m}_1$ and following the slope of $\bm{m}_1^{\prime}$ in positive $x$-direction for a distance of $\frac{\Delta_x}{2}$ and then the slope $\bm{m}_2^{\prime}$ for a distance of $\frac{\Delta_x}{2}$.

Inserting $\bm{m}_2^{\prime}$ from (\ref{eq:interface_jump}) and rearranging for $\bm{m}_1^{\prime}$ yields
\begin{equation}
\bm{m}_1^{\prime} = \big( \bm{m}_2 - \bm{m}_1 \big) \frac{2}{\Delta_x}\frac{A_2}{A_1 + A_2}
\end{equation}
With this expression for the derivative, we can introduce a virtual magnetization $\bm{m}_2^*$
\begin{equation}
\bm{m}_2^* = \bm{m}_1 + \Delta_x \bm{m}_1^{\prime} = \bm{m}_1 + \frac{2 A_2}{A_1 + A_2} \big( \bm{m}_2 - \bm{m}_1 \big),
\end{equation}
which then can be used in the three-point stencil centered around $\bm{m}_1$
\begin{equation}
\bm{m}^{\prime\prime} \approx \frac{\bm{m}_0 - 2 \bm{m}_1 + \bm{m}_2^*}{\Delta_x^2}
= \frac{\bm{m}_0 - \bm{m}_1 + \frac{2 A_2}{A_1 + A_2}\big( \bm{m}_2 - \bm{m}_1 \big)}{\Delta_x^2}
.
\end{equation}
The exchange field at the position of $\bm{m}_1$ then reads
\begin{equation}
\bm{H}_1^{\text{ex}}
= \frac{2 A_1}{\mu_0 M_{\text{s}, 1}} \frac{\bm{m}_0 - \bm{m}_1 + \frac{2 A_2}{A_1 + A_2}\big( \bm{m}_2 - \bm{m}_1 \big)}{\Delta_x^2}
.
\end{equation}

We can generalize this special case of a one-dimensional two-phase magnet to the most general one-dimensional case by assuming a cell-dependent exchange constant $A_i$ that varies between each cell $i$.
The effective field at site $i$ is then given as
\begin{equation}
\bm{H}_i^{\text{ex}}
=\frac{2 A_i}{\mu_0 M_{\text{s},i}} \frac{2}{\Delta_x^2} \Big[A_{i+1} \frac{\bm{m}_{i+1} - \bm{m}_i}{A_{i+1} + A_i} + A_{i-1} \frac{\bm{m}_{i-1} - \bm{m}_i}{A_{i-1} + A_i} \Big]
\end{equation}

When we consider a three-dimensional finite-difference grid with a regular cuboid discretization of $\Delta_\text{x}$, $\Delta_\text{y}$, and $\Delta_\text{z}$ along each axis, the exchange field becomes
\begin{align}
\begin{split}
\bm{H_i}^{\text{ex}}
&= \frac{2 A_{\bm{i}}}{\mu_0 M_{\text{s},\bm{i}}} \sum_{k=x,y,z} \frac{2}{\Delta_k^2} \Big[
A_{\bm{i} + \bm{e}_k} \frac{\bm{m}_{\bm{i} + \bm{e}_k} - \bm{m_i}  }{A_{\bm{i} + \bm{e}_k} + A_{\bm{i}}}
+ A_{\bm{i} - \bm{e}_k} \frac{\bm{m}_{\bm{i} - \bm{e}_k} - \bm{m_i}  }{A_{\bm{i} - \bm{e}_k} + A_{\bm{i}}}
\Big]
,
\end{split}
\end{align}
where $\bm{i}=\{i, j, k\}$ again is the three-dimensional cell index.
The boundary condition in equation (\ref{eq:m_x_2Admdn}) is properly accounted for by setting $A_{\bm{i}} = 0$ if the cell $\bm{i}$ is either non-magnetic or outside of the magnetic domain.

\subsection{Uniaxial anisotropy and external field in finite-differences}
The finite-difference discretization can be applied straightforwardly to the uniaxial anisotropy field and the external field.
Assuming a cell-wise definition of the material parameters, the effective field contribution of the uniaxial anisotropy in equation (\ref{eq:ani_micro}) becomes
\begin{equation}
\bm{H_i}^{\text{ani}} = \frac{2 K_{\bm{i}}}{\mu_0 M_{\text{s},\bm{i}}} \bm{e}_\text{u} (\bm{e}_\text{u} \cdot \bm{m_i})
,
\end{equation}
and the external field is applied cell-wise, that is, $\bm{H_i}^{\text{ext}}$.

\section{Exchange field in finite-elements}
Material jumps in finite-elements can be rigorously treated within the finite-element framework itself.
By calculating the effective field directly from the variation of the energy, all occurring boundary conditions are correctly taken into account \cite{abert_micromagnetics_2019}.
In the finite-elements discretization, it is a common choice to assume all material parameters to be piece-wise constant within each finite-element.
This property can be embedded directly in the framework by using a zeroth-order discontinuous Galerkin function space.
As we assume the magnetization to be a continuous field,
Moreover, the micromagnetic continuum approximation for the magnetic field can be directly represented by a linear continuous Galerkin function space.

In order to evaluate the effective field in finite-elements, we apply a set of test functions $\bm{v}$ onto equation (\ref{eq:Heff_via_dE})
\begin{equation}\label{eq:Heff_FEM}
\int_\Omega - \mu_0 M_\text{s} \bm{H}^\text{eff} \cdot \bm{v} \ \text{d}\bm{x} = \int_\Omega \frac{\delta E}{\delta \bm{m}} \cdot \bm{v} \ \text{d}\bm{x}
\end{equation}
This expression correctly treats jumps in material parameters, including $M_\text{s}$.

We highlight that the expression $\mu_0 M_\text{s}$ stands on the left-hand side of equation (\ref{eq:Heff_FEM}), not on the right-hand side.
When the spontaneous magnetization $M_\text{s}$ is inhomogeneous,
this subtle difference has a severe consequence for the evaluation of effective field terms which involve derivatives of the magnetization, such as the exchange field.
Would we have written
\begin{equation}
\begin{split}
\int_\Omega \bm{H}^{\text{ex}} \cdot \bm{v} \ \text{d}\bm{x}
& = \int_\Omega  \frac{1}{\mu_0 M_\text{s}} (2 \grad{\cdot A \grad{\bm{m}}})  \cdot \bm{v} \ \text{d}\bm{x}
\\ & = - \int_\Omega \frac{1}{\mu_0 } 2 A \grad{\bm{m}} \cdot \grad{\Big(\frac{1}{M_\text{s}} \bm{v}\Big)} \ \text{d}\bm{x}
+ \int_{\partial\Omega} \frac{1}{\mu_0 M_\text{s}} 2 A \grad{\bm{m}} \cdot \bm{n} \cdot \bm{v} \ \text{d}\bm{s}
\end{split}
\end{equation}
when evaluating the exchange field, the integration by parts would introduce new boundary terms which would require explicit handling.

In order to avoid such potential errors resulting from discontinuities in material parameters, we introduce a standard problem that applies to both finite-difference and finite-element micromagnetic solvers.

\section{Proposed standard problem}
In the following, we propose a standard problem calculating the coercive field for the unpinning of a domain wall from the interface of a two-phase magnetic rod.
The considered micromagnetic energy contributions are the exchange interaction, uniaxial anisotropy, and an external field.
The proposed problem has a simple analytic solution when neglecting demagnetization, making it well suited for testing micromagnetic simulation software.

We consider a domain wall pinned at the interface between a soft magnetic and a hard magnetic phase.
Given a higher anisotropy or exchange interaction in the hard magnetic phase, the micromagnetic energy stored in the domain wall is higher in the hard magnetic phase than in the soft magnetic phase.
As a result, the domain wall must overcome this energy difference when pushed from the soft to the hard magnetic phase by an external field.
This effect is referred to as domain wall pinning at a phase boundary.

The applied external field at which the magnetic domain detaches from the phase boundary is called the pinning field $H_\text{p}$.
An analytical expression for the pinning field can be derived in the micromagnetic model for a sharp phase boundary, as is described in \cite{kronmuller_micromagnetic_2002}.
Using the spontaneous polarization $J_\text{s} = \mu_0 M_\text{s}$ and
assuming micromagnetic parameters $A^\text{I}, K^\text{I}, J_\text{s}^\text{I}$ in the first magnetic phase and $A^\text{II}, K^\text{II}, J_\text{s}^\text{II}$ in the second magnetic phase with ratios $\epsilon_\textit{A}=A^\text{I}/A^\text{II}$, $\epsilon_\textit{K}=K^\text{I}/K^\text{II}$, and $\epsilon_\textit{J}=J_\text{s}^\text{I}/J_\text{s}^\text{II}$, the analytical formula for the pinning field $H_\text{p}$ reads \cite[eq. 10]{kronmuller_micromagnetic_2002}
\begin{equation}
	H_\text{p} = \frac{2 K^\text{II}}{J_\text{s}^\text{II}} \frac{1-\epsilon_\textit{K} \epsilon_\textit{A}}{\Big({1+\sqrt{\epsilon_\textit{J} \epsilon_\textit{A}}}\Big)^2}
	.
\end{equation}

As the pinning field depends on all three material parameters ratios, this problem is sensitive to the correct treatment of discontinuities in the exchange, uniaxial anisotropy, and spontaneous magnetization parameters.
Furthermore, we can conveniently vary and test individual material jumps, as well as combinations thereof.
Combined with the fact that this problem has an analytical solution, the numerical calculation of the pinning field is well suited as a standard problem to test interface conditions in micromagnetic codes.

\subsection{Problem specification}
\begin{figure}[h!]
	\centering
	\includegraphics{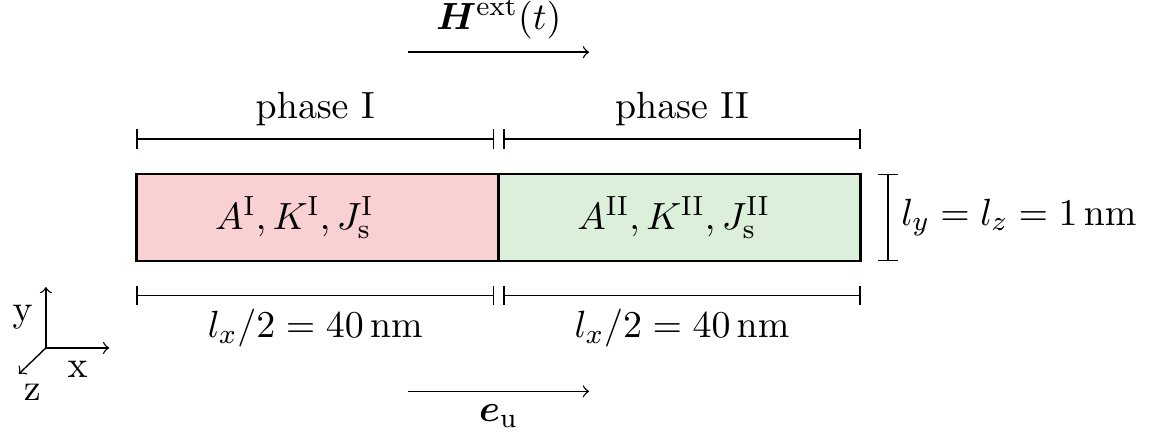}
	\caption{Illustration of the two-phase magnetic rod geometry (not to scale) with material parameters $A, K, J_\text{s}$ for each phase and indication for the direction of the applied external field $\bm{H}^{\text{ext}}(t)$ and the uniaxial anisotropy vector $\bm{e}_{\text{u}}$.}
	\label{fig:sp_layout}
\end{figure}

We consider a rod in the shape of a cuboid consisting of two magnetic regions with a sharp phase boundary, as illustrated in Figure \ref{fig:sp_layout}.
The dimensions of each magnetic phase are $l_x = \SI{40}{nm}$ and $l_y = l_z = \SI{1}{nm}$.
We use two sets of micromagnetic parameters, describing a soft magnetic and a hard magnetic material.

$$A_\text{soft} = \SI{0.25e-11}{J/m}, \ A_\text{hard} = \SI{1e-11}{J/m}$$
$$ K_\text{soft} = \SI{1e5}{J/m^3}, \ K_\text{hard} = \SI{1e6}{J/m^3}$$
$$ J_{\text{s,soft}} = \SI{0.25}{T}, \ J_{\text{s,hard}} = \SI{1.00}{T}$$

We always use the hard magnetic parameters for phase II, i.e. $A^\text{II} = A_\text{hard}, K^\text{II} = K_\text{hard}, J_\text{s}^\text{II} = J_{\text{s,hard}}$.
In phase I, we set the material parameter to either the soft magnetic or the hard magnetic value.
So, for example, $A^\text{I}$ is either $A_\text{soft}$ or $A_\text{hard}$.
This allows us to test individual material jumps as well as combinations thereof.

In both regions, the uniaxial anisotropy unit vector points in positive $x$-direction and the damping parameter is set to one.
$$\bm{e}_{\text{u}} = (1, 0, 0)$$
$$ \alpha = 1$$

The initial magnetization is assumed to point approximately in positive $x$-direction in phase I, and in negative $x$-direction in phase II, separated by a domain wall at the interface.
The domain wall is pinned at the phase boundary by an external field applied in positive $x$-direction.
The field increases linearly as a function of the simulation time $t$ in seconds with a certain field rate $r$ in $\SI{}{A/m/s}$.
$$\bm{H}^{\text{ext}}(t) = ( r * t, 0, 0)$$

When the external field becomes larger that the pinning field $H_\text{p}$, the domain wall unpins from the interface and propagates through phase II, ultimately being pushed out of the magnetic rod.
The aim of this standard problem is to calculate this pinning field using micromagnetic solvers.

\section{Numerical Results}
In this section, we calculate the pinning field using dynamic micromagnetic simulations and compare the results to the analytical solution.
As an initial magnetization, we set the magnetization in phase I to $(1.0, 0.3, 0)/\norm{(1.0, 0.3, 0)}\approx(0.958, 0.287, 0)$, such that it is facing in positive $x$-direction with a tilt in the direction of the $y$-axis.
For phase II, we set the magnetization to $(-1.0, 0.3, 0)/\norm{(-1.0, 0.3, 0)}$.
This initial configuration is easy to configure in simulation tools and evolves into a domain wall when integrated in time.
The tilt in the $y$-direction is used to avoid numerical problems due to symmetry.
The system is evolved in time by integrating the LLG equation.
The applied field is increased linearly with a rate of $r = \SI{2e7}{T/s} = \SI{1.59154943e6}{A/m}$.
We define the numerically calculated pinning filed $H_\text{p}^\textit{num}$ as the point, where the average $x$-component of the magnetization $\langle m_x \rangle$ becomes larger than $(1 - \SI{1e-3}{}) = 0.999$.
This definition of the pining field is straightforward to implement in micromagnetic solvers.
We are aware that this definition leads to slightly higher pinning fields, as it includes the time the domain wall needs to propagate through phase II.
More precise definitions could be used, such as the point where the domain wall just crossed the interface, but we will use the above definition throughout this manuscript for the sake of simplicity.

\begin{figure}[h!]
	\centering
	\resizebox{\linewidth}{!}{
		\includegraphics{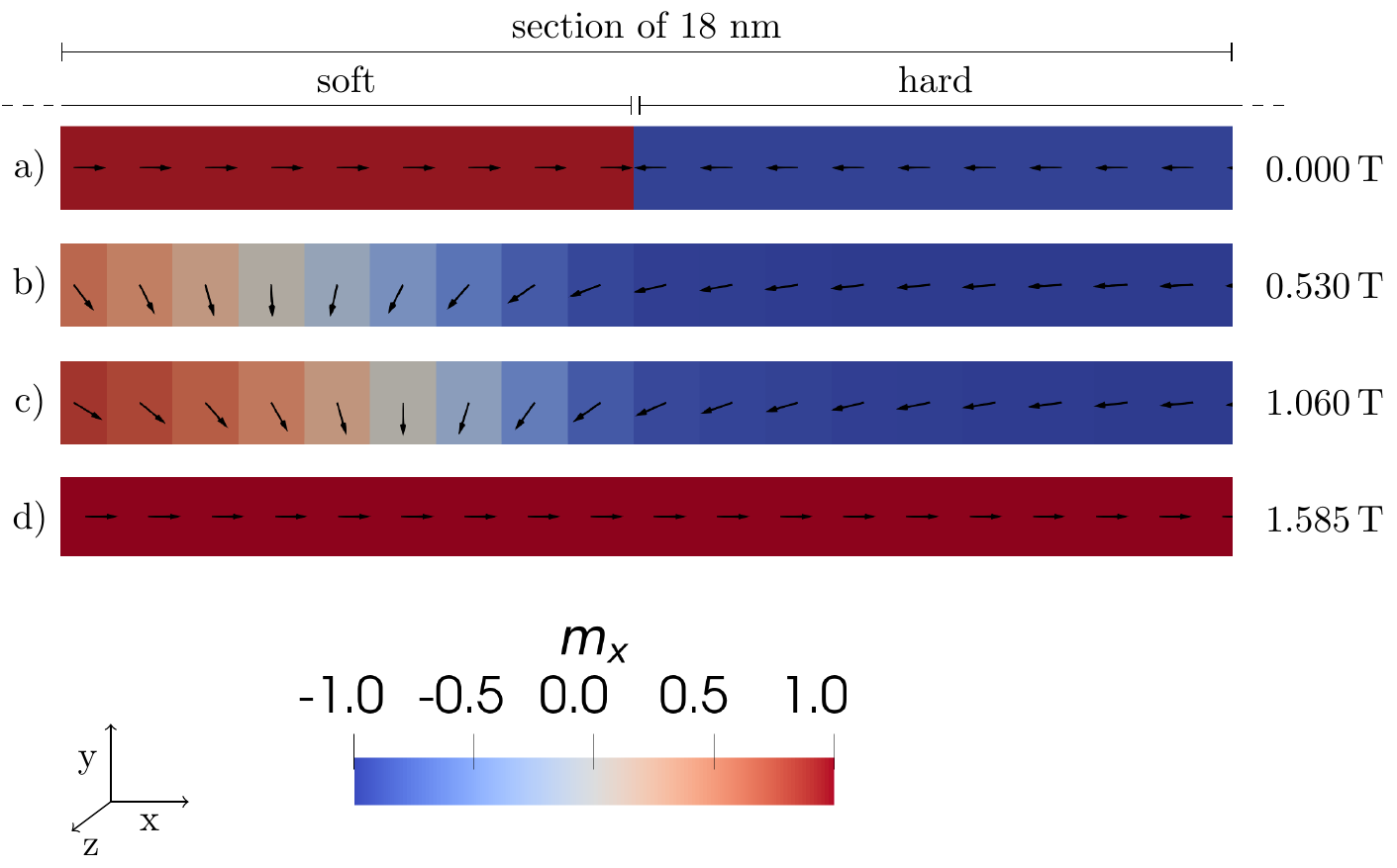}
	}
	\caption{
		Section of the magnetic rod at the interface between the soft and hard magnetic regions during external field increase from $\mu_0 H_x = \SI{0}{T}$ to $\SI{2}{T}$ for four selected values of the external field.
		The color code indicates the $x$-component of the magnetic field.
		a) Initial magnetization, b) pinned domain wall, c) compressed domain wall, d) domain wall unpinned and propagated through the hard magnetic region.
	}
	\label{fig:mag_stripes_section}
\end{figure}

For the case that all three material parameters jump, the analytical pinning field evaluates to $\mu_0 H_\text{p} \approx \SI{1.568}{T}$.
Figure \ref{fig:mag_stripes_section} shows a section of the magnetic rod centered at the phase transition as obtained by the finite-difference code \textit{magnum.af} \cite{heistracher_hybrid_2020}. The arrows indicate the magnetization direction, and the color encodes the $x$-component of the magnetization.
At an external field of \SI{1.585}{T}, the criterion for $H_\text{p}^\textit{num}$ is met as the domain wall was unpinned and pushed out of the rod, leaving a nearly homogeneous magnetization pointing in the positive $x$-direction.
This result is in good agreement with the expected pinning field of \SI{1.568}{T}, considering the additional time the domain wall needs to propagate through the hard magnetic region.

\subsection{Comparison of micromagnetic solvers}
In the following, we use this domain wall pinning field to compare several micromagnetic codes, namely the finite-difference codes \textit{OOMMF} \cite{m._j._donahue_oommf_1999}, \textit{mumax3} \cite{vansteenkiste_design_2014} in versions 3.9.3 and 3.10, and \textit{magnum.af}, as well as the finite-element code \textit{magnum.fe} \cite{abert_magnumfe_2013}.
The input scripts used to obtain the presented results can be found in the supplementary material for each respective simulation tool.

\subsubsection{Jump in $A$, $K$, $J_\text{s}$}\label{sec:AJK}
\begin{figure}[h!]
	\centering
	\includegraphics{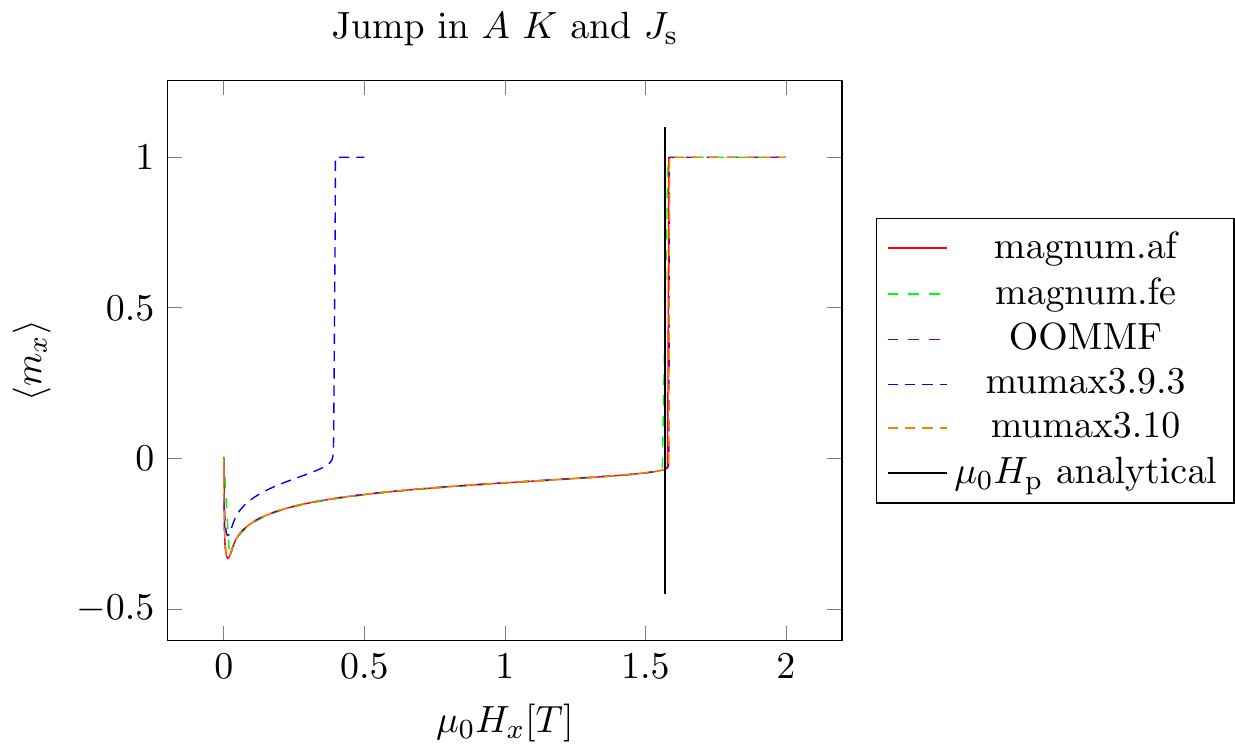}
	\caption{Comparison of different micromagnetic solvers calculating the domain wall pinning field.
		Plotted is the average $x$-component of the magnetization field over the applied external field.
		The analytical pinning field is indicated by the black vertical line.}
	\label{fig:jump_in_jak}
\end{figure}

We start with the case that all material parameters jump, that is, $A^\text{I}=A_\text{soft}, K^\text{I}=K_\text{soft}, J_\text{s}^\text{I}=J_{\text{s,soft}}$.
Figure \ref{fig:jump_in_jak} shows the average $x$-component of the magnetization plotted over the external field strength.
All codes yield results close to the expected analytical pinning field, except for \textit{mumax3.9.3}.
Using version 3.9.3, we observe a substantial deviation from the expected pinning field, predicting a value of around \SI{0.397}{T}.
This error is introduced by taking a harmonic mean not only over the exchange constant but over the spontaneous magnetization parameter as well, as seen in \cite[eq. 9]{vansteenkiste_design_2014}.
Fortunately, this error is corrected in the current version 3.10 and is now consistent with the other simulation tools.

With this standard problem, not only the case that all material parameters jump can be examined, but also jumps in individual parameters or combinations thereof.
This property makes it possible to precisely trace potential implementation errors by consecutively testing different combinations of material jumps, as we demonstrate in the following by the example of \textit{mumax3.9.3}.

\subsubsection{Jump in $A$}
\begin{figure}[h!]
	\centering
	\includegraphics{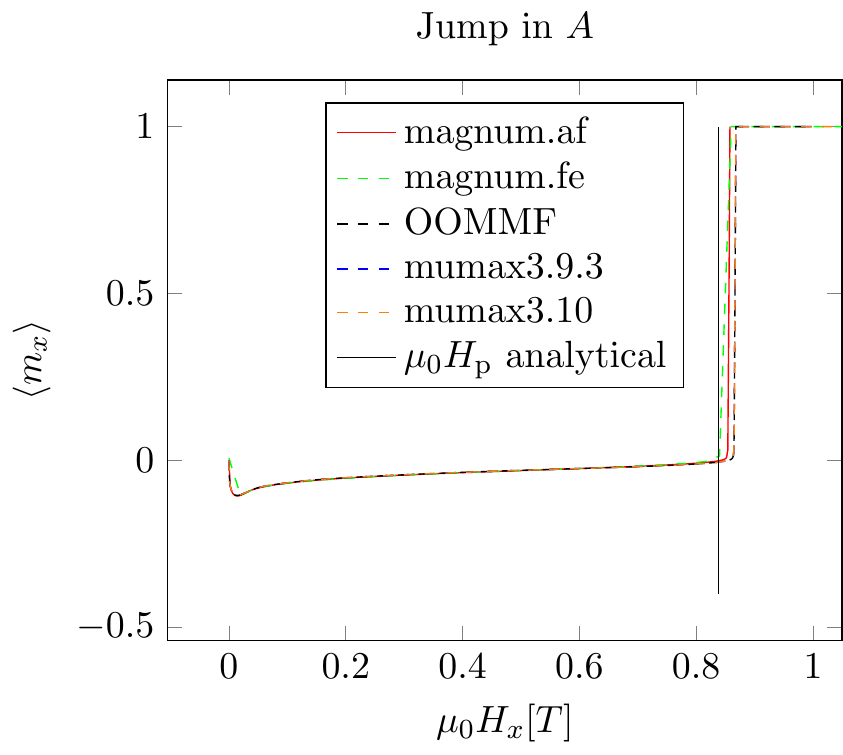}
	\caption{Comparison of micromagnetic codes calculating the pinning field for the case that only the exchange constant $A$ jumps between phase I and phase II of the magnetic rod.
	}
	\label{fig:jump_in_a}
\end{figure}

In the following example, we discuss the case that only the exchange constant $A$ varies between the two phases, i.e., $A^\text{I} = A_\text{soft}$ and $A^\text{II} = A_\text{hard}$. 
The uniaxial anisotropy and spontaneous magnetization parameters are set to the hard magnetic values, $K^\text{I} = K^\text{II} = K_\text{hard}$ and $J_\text{s}^\text{I} = J_\text{s}^\text{II} = J_{\text{s,hard}}$.
For these values, the analytical pinning field evaluates to $ \mu_0 H_\text{p} \approx \SI{0.838}{T}$.
Figure \ref{fig:jump_in_a} compares the calculated pining fields for the jump in $A$.	
In this case, all codes yield the predicted results, including \textit{mumax3.9.3}.
This agreement indicates that the implementation of discontinuities in the exchange parameter does not cause the previous discrepancy.
Rather, the error might originate in the treatment of jumps in parameters for the uniaxial anisotropy or the spontaneous magnetization.

\subsubsection{Jump in $A$ and $J_\text{s}$}
To trace the error further, we investigate the case where both the exchange constant $A$ and the spontaneous polarization $J_\text{s}$ exhibit a jump at the interface.
The micromagnetic parameters therefore are $A^\text{I}=A_\text{soft}$, $A^\text{II}=A_\text{hard}$, $J_\text{s}^\text{I}=J_{\text{s,soft}}$, $J_\text{s}^\text{II}=J_{\text{s,hard}}$, and $K^\text{I}=K^\text{II}=K_\text{hard}$.
For these values, the analytical pinning field evaluates to $ \mu_0 H_\text{p} \approx \SI{1.206}{T}$.

\begin{figure}[h!]
	\centering
	\includegraphics{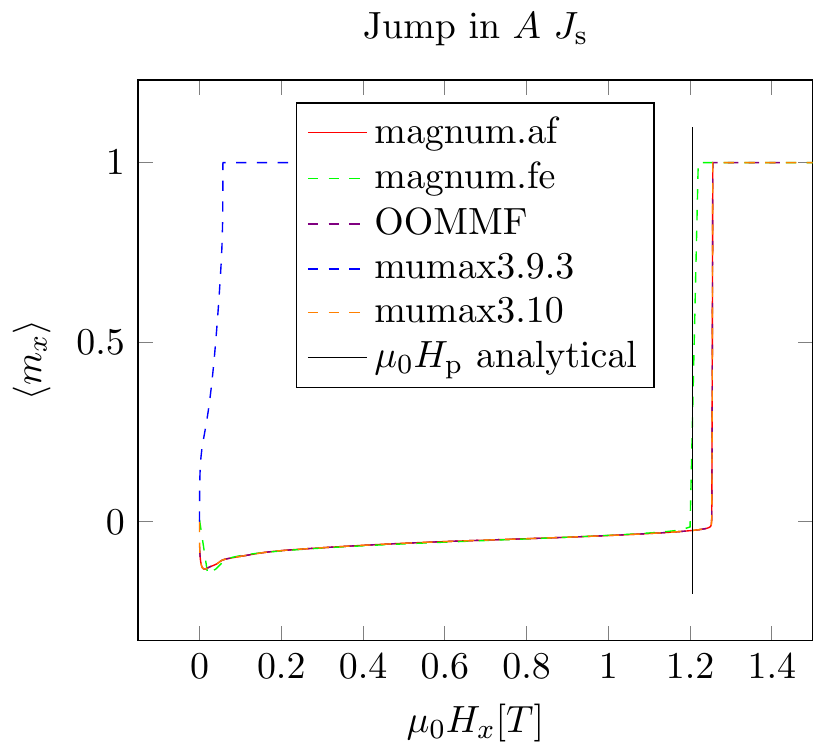}
	\caption{Comparison of micromagnetic codes calculating the pinning field for the case that both the exchange constant $A$ and the spontaneous polarization $J_\text{s}$ jump between phase I and phase II.
	}
	\label{fig:jump_in_a_ms}
\end{figure}

Figure \ref{fig:jump_in_a_ms} compares the results of the simulation codes for a jump in $A$ and $J_\text{s}$.
We observe that all codes yield the expected pinning field, again except for \textit{mumax3.9.3}.
In combination with the previous two cases, this result indicates that jumps in the spontaneous polarization $J_\text{s}$ are handled incorrectly in the implementation of \textit{mumax3.9.3}.
The ability to selectively vary jumps in material parameters is of high value when tracing errors during the development of micromagnetic simulation tools.

\subsection{All combinations of material jumps}
In the following, we present an overview of all possible combinations of material jumps when individually varying the micromagnetic parameters in phase I.
We denote a jump in the micromagnetic exchange constant $A$, i.e., $A^I=A_\text{soft}$ and $A^{II}=A_\text{hard}$, with $a$.
Likewise, we refer to a jump in $K$ with $k$ and a jump in $J_\text{s}$ with $j$.
Then all possible combinations of material jumps can be classified into $2^3 = 8$ cases, denoted as \{$akj,ak,aj,a,kj,k,j,$" "\}, where " " refers to the case with no jump of any material parameter.

Table \ref{tab:overview} compares the analytical pinning field $H_\text{p}$ with the calculated pinning field $H_\text{p}^\textit{num}$ for all combinations of jumps in material parameters $A$, $K$, and $J_\text{s}$.
The tick denotes that the respective parameter varies between the phases, whereas, for a missing tick, the hard magnetic parameter is used.
We compare the analytical solution with the numerical results obtained by the finite-difference solver \textit{magnum.af} and observe good agreement.
The offset in the order of tens of \SI{}{mT} is caused by the relaxation process using integration in the time domain and depends on the rate of the external field increase.
Reducing the rate decreases the offset while increasing simulation time.
Table \ref{tab:overview} is intended to be used as a reference for testing this standard problem in other micromagnetic codes.

\begin{table}[h!]
	\centering
	\begin{tabular}{|c|c|c|c|l|l|}
		\hline
		case & $A$ jump & $K$ jump & $J_\text{s}$ jump & $H_\text{p}$[\SI{}{T}]  & $H_\text{p}^\textit{num}$[\SI{}{T}] \\ \hline
		$akj$& \texttimes   & \texttimes   & \texttimes     & $\SI{1.568}{}$ & $\SI{1.585}{}$     \\ \hline
		$ak$ & \texttimes   & \texttimes   &                & $\SI{1.089}{}$ & $\SI{1.116}{}$     \\ \hline
		$aj$ & \texttimes   &              & \texttimes     & $\SI{1.206}{}$ & $\SI{1.256}{}$     \\ \hline
		$a$  & \texttimes   &              &                & $\SI{0.838}{}$ & $\SI{0.868}{}$     \\ \hline
		$kj$ &              & \texttimes   & \texttimes     & $\SI{1.005}{}$ & $\SI{1.020}{}$     \\ \hline
		$k$  &              & \texttimes   &                & $\SI{0.565}{}$ & $\SI{0.582}{}$     \\ \hline
		$j$  &              &              & \texttimes     & $\SI{0.0}{}$   & $\SI{0.068}{}$     \\ \hline
		" "  &              &              &                & $\SI{0.0}{}$   & $\SI{0.068}{}$     \\ \hline
	\end{tabular}
	\caption{Table showing all possible combinations of jumps in $A$, $K$, and $J_\text{s}$ with according analytical pining field $H_\text{p}$ and numerically calculated field $H_\text{p}^\textit{num}$ as obtained by the simulation tool \textit{magnum.af} with a an external field rate of $r=\SI{2e7}{T/s}$.}
	\label{tab:overview}
\end{table}

\section{Conclusion}
We preset a standard problem that calculates the coercive field for the unpinning of a domain wall from a two-phase magnetic rod interface.
This problem is sensitive to discontinuities in the micromagnetic parameters for the exchange interaction, the uniaxial anisotropy, and the spontaneous magnetization.
We derive the interface condition for the exchange interaction in finite-differences and verify that our simulations agree with the analytical solutions.
The proposed problem is well suited to verify the correct implementation of material jumps in both finite-difference and finite-element micromagnetic simulation codes.
Therefore, we encourage authors of micromagnetic simulation tools to use this domain wall pinning problem to test their solvers for proper treatment of discontinuities in material parameters.

\subsection*{Acknowledgments}
The financial support by the Austrian Federal Ministry for Digital and Economic Affairs, the National Foundation for Research, Technology and Development and the Christian Doppler Research Association is gratefully acknowledged.

\subsection*{Author Contributions}
\textbf{Paul Heistracher:} Methodology, Software, Validation, Writing - Original Draft.
\textbf{Claas Abert:} Conceptualization, Methodology, Writing - Review \& Editing.
\textbf{Florian Bruckner:} Methodology, Writing - Review \& Editing.
\textbf{Thomas Schrefl:} Conceptualization, Writing - Review \& Editing.
\textbf{Dieter Suess:} Conceptualization, Supervision, Funding acquisition, Writing - Review \& Editing.

\subsection*{Competing Interests}
The authors declare no competing financial interests.

\subsection*{Additional Information}
Input scripts for all used simulation tools are found in the supplementary material.

\bibliographystyle{ieeetr}
\bibliography{ref}

\end{document}